\documentstyle[12pt,epsf,cite]{article}
\def\g{\gamma}
\def\m{\mu^2}
\def\G{\Gamma}

\def\i{\prime}
\def\am{(\alpha_1-\beta_1)}
\def\ap{(\alpha_1+\beta_1)}
\def\amn{\am_{\pi^0}}
\def\amc{\am_{\pi^{\pm}}}

\def\gg{\g \g\to \pi^0 \pi^0}

\def\mpp{M_{++}}
\def\mm{M_{+-}}
\def\sp{s'}
\def\tp{t'}
\def\be{\begin{equation}}
\def\ee{\end{equation}}
\def\beq{\begin{eqnarray}}
\def\eeq{\end{eqnarray}}
\begin{document}
\begin{center}
{\large \bf Comments to the article "The polarizability of the pion: no conflict 
between dispersion theory and chiral perturbation theory"}

\vspace{0.5cm}

{\bf L.V. Fil'kov and V.L. Kashevarov}
\vspace{0.5cm}

{\it Lebedev Physical Institute, Leninsky Prospect 53,
Moscow 119991, Russia.\\
E-mail: filkov@sci.lebedev.ru}
\end{center}
\vspace{0.5cm}

\begin{center}
Abstract
\end{center}
\vspace{0.2cm}

The statement of the authors of the article \cite{pasq}, that
spurious singularities occur in the dispersion relation approach of
the works \cite{rad,fil1,fil2,fil3}, is analyzed. It has been shown
that there are not any additional singularities in this approach 
and the disagreement between the predictions of the dispersion relations 
and ChPT for $\amc$ remains. Reasons for the negative results
of the description of the process $\gg$ in \cite{pasq} are also discused.
\vspace{0.3cm}


\vspace{1cm}

In the paper of B. Pasquini, D. Drechesel, and S. Scherer \cite{pasq} 
the authors attempt to eliminate discrepancies between predictions of the 
dispersion relations (DR)
in \cite{rad,fil1,fil2,fil3} and the results obtained in the framework of 
chiral perturbation theory (ChPT) \cite{burgi,gas1, gas2}.
The authors claimed that, as the absorptive part of the Compton amplitudes 
in \cite{fil1,fil2,fil3} is expressed by Breit-Wigner poles with coupling 
constants and decay widths dependent on energy, there must appear additional
spurious singularities. As a result, the values of the polarizabilities obtained
in \cite{fil1,fil2,fil3} have to be modified essentially.

The process $\g\g\to\pi\pi$ is described by the following invariant variables:
\be
t=(k_1+k_2)^2, \quad  s=(q_1-k_1)^2, \quad  u=(q_1-k_2)^2,
\ee
where $q_1(q_2)$ and $k_1(k_2)$ are the pion and photon four-momenta, 
respectively.

In order to analyze this process, DRs at fixed $t$ with one subtraction
at $s=\m$ (where $\mu$ is the pion mass) have been constructed 
\cite{fil2,fil3} for the helicity amplitudes $\mpp$ and $\mm$ of this process.
Via the cross symmetry these DRs are identical to DRs with two subtractions.
The subtraction functions were determined with the help of DRs at fixed
$s=\mu^2$ with two subtractions and the subtraction constants were expressed 
through the sum and the difference of
the electric and magnetic dipole and quadrupole pion polarizabilities. 
It is worth to note that these DRs do not have any expansions and so 
they can be used for the determination of the polarizabilities in the 
region of both low and intermediate energies.

Besides, dispersion sum rules (DSR) have been constructed \cite{fil1,fil2} for
the dipole and quadru\-pole pion polarizabilities using DRs at fixed $u=\mu^2$ 
without and with one subtraction, respectively. The DSR for the difference of 
the electric and magnetic dipole polarizabilities reads
\be
\am=
\frac{1}{2\pi^2\mu}\left\{\int\limits_{4\m}^{\infty}~\frac{
Im\mpp(\tp,u=\m)~d\tp}{\tp}+\int\limits_{4\m}^{\infty}~\frac{
Im\mpp(\sp,u=\m)~d\sp}{\sp-\m}\right\}.
\label{dsr1m}
\ee
  
The imaginary parts of the amplitudes in these DRs and DSRs are saturated 
by the contributions  of meson resonances by using Breit-Wigner expressions. 
It should be noted that these expressions are used to calculate the 
imaginary parts of 
the amplitude only. For example, the contribution of the vector and
axial-vector mesons are calculated with the help of the expression
\be
Im\mpp^{(V)}(s,t)=\mp s~Im\mm^{(V)}(s,t)=\pm4g_V^2 s\frac{\G_0}
{(M_V^2-s)^2+\G_0^2},
\ee
where $M_V$ is the vector meson mass, the sign "+" corresponds to the
 contribution of the $a_1$ and $b_1$ mesons and
$$
g_V^2=6\pi\sqrt{\frac{M_V^2}{s}}\left(\frac{M_V}{M_V^2-\m}\right)^3
\G_{V\to\g\pi},
$$
\be
\G_0=\left(\frac{s-4\m}{M_V^2-4\m}\right)^{\frac32}M_V\G_V.
\ee
Here $\G_V$ and $\G_{V\to \g\pi}$ are the full width and the decay width into
$\g\pi$ of these mesons, respectively. A dependence of the width on 
the energy is
conditioned by the threshold behaviour. The energy dependence of the constant
$g_V$ appears via an expression of the total cross section of the process
$\g\pi\to\g\pi$ through the vector meson contribution. Moreover, a similar
dependence of the imaginary part of the amplitude on $1/\sqrt{s}$ is caused by
the unitarity condition and does not lead to any additional singularity.

In order to check the possibility of the appearance of additional spurious 
singularities,
we calculated the contribution of all mesons, except $\sigma$, to the 
DSR~(\ref{dsr1m}) in the zero-width approximation. 
The results of these calculations ($\am_{z-w}$ and $\ap_{z-w}$)
are listed in Table 1 and Table 2 together with the full calculations 
obtained in Ref. \cite{fil2}.

\begin{table}
\caption{The DSR predictions for the dipole polarizabilities of the charged
pions in units of $10^{-4}$\,fm$^3$.
} 
\centering
\begin{tabular}{ccccccccccc}\hline
 &$\rho$&$b_1$&$a_1$&$a_2$&$f_0$&$f_0^{\i}$&$\sigma$&$\Sigma$
&$\Delta\Sigma$ \\ \hline
$\am_{full}$&-1.15&0.93 &2.26 &1.51 &0.58&0.02&9.45&13.60&2.15 \\ \hline
$\am_{z-w} $&-1.11&0.85 &3.39 &1.51 &0.59&0.03&9.45&13.70&     \\ \hline
$\ap_{full}$&0.063&0.021&0.051&0.031& -  & -  & -  &0.166&0.024 \\ \hline
$\ap_{z-w} $&0.072&0.022&0.062&0.032& -  & -  & -  &0.188&      \\ \hline
\end{tabular}
\end{table}

\begin{table}
\caption{The DSR predictions for the dipole polarizabilities of the $\pi^0$
meson.}
\centering
\begin{tabular}{cccccccccc}\hline
 &$\rho$&$\omega$&$\phi$&$f_0$&$f_0^{\i}$&$\sigma$&$\Sigma$
&$\Delta\Sigma$ \\ \hline
$\am_{full}$&-1.58 &-12.56&-0.04 &0.60 &0.02 &10.07&-3.49  &2.13 \\ \hline
$\am_{z-w} $&-1.99 &-11.81&-0.04 &0.61 &0.02 &10.07&-3.14  &     \\ \hline
$\ap_{full}$&0.080 &0.721 & 0.001 &  -  &  - &  -  & 0.802 &0.035 \\ \hline
$\ap_{z-w} $&0.122 &0.703 & 0.001 &  -  &  - &  -  & 0.826 &      \\ \hline
\end{tabular}
\end{table}

As was noted in \cite{fil1}, we consider the $\sigma$ meson as an
effective description of the strong $S$-wave $\pi\pi$ interaction using the 
broad Breit-Wigner resonance expression for the imaginary part of the 
amplitude. The parameters of such a sigma meson were found from the fit to 
the experimental data \cite{mars} for the process $\gg$ in the energy 
region of $\sqrt{t}=270\div 825$ MeV. Therefore, such a "$\sigma$ meson" 
describes the contribution of the full $S$-wave $\pi\pi$ interaction to the 
process $\gg$ and its decay width could be bigger than for a real $\sigma$ 
meson. If we use energy independent values of the decay width and the 
coupling constant of the $\sigma$ meson, then the results of the 
calculations are not changed essentially.
For example, the value of $\am_{\pi^{\pm}(z-v)}$ would be equal to 13.1.

As seen from these Tables the results of the calculations in the zero-width 
approximation practically coincide with the calculations performed in 
Ref. \cite{fil2} beyond such an approximation. 
This result is evidence of the absence of any additional spurious 
singularities in the approach which was used in \cite{fil1,fil2,fil3}.

It is worth noting that the calculation of $\amc$ with the help of the DSR 
at finite
energy \cite{petr}, which takes into account the $s$-channel contributions and
Regge asymptotic only,
yielded $\amc=10.3\pm 1.3$. This value practically coincides
with the result of the calculation in \cite{fil2} (see Table 1) and also confirms 
the absence of additional singularities in the works under consideration. 

By the way, if one follows the statement of the authors of \cite{pasq},
spurious singularities could also appear in their work \cite{drec} via
the kinematical coefficients, for example, in the expressions (40) 
for $Im_{t}A_i(0,t)$.           	

The authors of the paper \cite{pasq} calculated  
the dipole pion polarizabilities by using the expressions:
\be
\label{omn}
\am=-\frac{1}{4\pi\mu}\left(A^V_1(\m,0)-\frac1\pi~\int_{4\m}^\infty~d\tp
\frac{H^I_{00}(\tp) Im\left[(\Omega^I_0)^{-1}(\tp)\right]}{\tp}\right),
\ee
\be
\ap=-\frac1{4\pi\m}\left(A^V_2(\m,0)+A_2^{f_2}(\m,0)\right).
\ee 
where $\Omega^I_J(t)$ is the Omn\`es function. The function $H^I_{00}(t)$
is expressed by means of the contribution of the vector mesons, which is 
assumed in the form
\be
\label{vect}
 A^V_1(s,t)=-2e^2 R_V\left[\frac{s}{s-M_V^2} +\frac{u}{u-M_V^2}\right],
\ee
\be
 A^V_2(s,t)=2e^2 R_V\left[\frac{1}{s-M_V^2} +\frac{1}{u-M_V^2}\right],
\ee
where
\be
R_V=\frac{24\pi M_V^3 \Gamma(V\to\pi\g)}{e^2~(M_V^2-\m)^3}.
\ee
However, Eq.(\ref{omn}) is a solution of the DR for the amplitude 
$A_1(s,t)$ in the work \cite{pasq}. 
Therefore, the contributions of the vector mesons should be taken into account
as poles in the $s$ and $u$ channels:
\be
\label{vect-d}                                            
 A^V_1(s,t)=-2e^2 R_V\left[\frac{M_V^2}{s-M_V^2} +\frac{M_V^2}{u-M_V^2}\right].
\ee

Use of the expression (\ref{vect}) in Eq.(\ref{omn}) in \cite{pasq} 
is the main reason of the unsatisfactory description of the process $\gg$ and the 
erroneous value of $\amn=6.62$ obtained in this work. This value differs from
the ChPT result $\amn=-1.90$ both in the quantity and in the sign.

In the work \cite{pasq} the contribution of the $S$-wave $\pi\pi$ interaction
is determined by integrating Eq.(\ref{omn}) and a corresponding expression
for the DRs 
from the threshold up to 800 MeV.
However, the $S$-wave contribution is not limited to this interval of energy
and  strongly depends on the upper integration limit $(\Lambda)$. 
As was shown in Ref.\cite{fil3}, the results of integration over $\tp$
in the DRs are not changed for $\Lambda$ greater than (5 GeV)$^2$ only.
Such an extension of the integration region will lead to an essential increase
of the contribution of the $S$-wave $\pi\pi$ interaction to $\am$.

For the reaction $\gg$ the Born term is equal to zero and the main contribution 
to this process in the energy region up to 800 MeV
is given by $S$-wave $\pi\pi$ interaction. Therefore, an analysis of 
this process in this energy region allows a determination of the parameters 
of the $S$-wave  with sufficient accuracy. 
The unsatisfactory description of this process in the energy region 
under consideration in the work \cite{pasq} is probably 
connected, in particular, with the 
incorrect determination of the $S$-wave contribution in this work.

In conclusion, we showed that there are not any additional spurious
singularities in the dispersion approach of \cite{rad,fil1,fil2,fil3}.
Unfortunately, the difference between the predictions of the DSRs \cite{fil2,petr}
and ChPT \cite{burgi,gas2} for $\amc$ remains. 
This discrepancy is connected with a
different account of the contribution of the $\sigma$ meson and the vector 
mesons
in the DSR and the ChPT calculations (see \cite{bonn}). In the present ChPT the
$\sigma$ meson is taken into account only partially through the two-loop
calculations. Moreover, the contributions of the vector mesons are
considered as poles at $s=M_V^2$ in DSRs and as the Born terms in ChPT. 
It results
in the difference of the predictions for $\mpp^{(V)}$ in these calculations 
by a factor $M_V^2/\m$.

The authors thank A.I. L'vov for usuful discussions.
This research is part of the EU integrated initiative hadron physics project
under contract number RII3-CT-2004-506078 and was supported in part by the
Russian Foundation for Basic Research (Grant No. 05-02-04014).


\begin{thebibliography}{99}
\bibitem{pasq} B. Pasquini, D. Drechsel, and S. Scherer, arXiv:0805.0213 
[hep-ph].
\bibitem{rad} L.V. Fil'kov, I. Guiasu and E.E. Radescu, Phys. Rev. D {\bf 26},
3146 (1982).
\bibitem{fil1} L.V. Fil'kov and V.L. Kashevarov, Eur.Phys.J. A {\bf 5}, 285
(1999).
\bibitem{fil2} L.V. Fil'kov and V.L. Kashevarov, Phys. Rev. C {\bf 72}, 035211
(2005).
\bibitem{fil3} L.V. Fil'kov and V.L. Kashevarov, Phys. Rev. C {\bf 73}, 035210
(2006).
\bibitem{burgi} U. B\"urgi, Nucl. Phys. B {\bf 479}, 392 (1997).
\bibitem{gas1} J. Gasser, M.A. Ivanov, and M.E. Sainio, Nucl. Phys. B {\bf 728},
31 (2005).
\bibitem{gas2} J. Gasser, M.A. Ivanov, and N.E. Sainio,
Nucl. Phys. B {\bf 745}, 84 (2006).
\bibitem{petr} V.A. Petrun'kin, Sov. J. Part. Nucl. D{\bf 12}, 278 (1981);
A.I. L'vov and V.A. Petrun'kin, Sov. Phys.-Lebedev Inst. Rep. {\bf 12}, 39 (1985).
\bibitem{mars} H. Marsiske, D. Antreasyan, H.W. Bartels {\em et al.}, Phys. Rev.
D {\bf 41}, 3324 (1990).
\bibitem{drec} D. Drechsel, {\em et al.}, Phys. Rev. C {\bf61}, 015204 (1999).
\bibitem{bonn} L.V. Fil'kov and V.L. Kashevarov, arXiv:0802.0965 (nucl-th);
Proceedings of "NSTAR 2007", Bonn, Germany, 05-08 September (2007).
\end{thebibliography}
\end{document}